\documentclass[traditabstract]{aa} 
\usepackage{amsmath}
\usepackage{txfonts} 
\usepackage{graphicx}
\usepackage{subfigure}
\usepackage{color}
\usepackage{bm}
\usepackage{natbib}
\usepackage{url}
\usepackage[]{natbib}
\usepackage{lscape}
\usepackage{longtable}
\usepackage{rotating}
\newcommand{\spitzer}{\mbox{\it Spitzer}}
\newcommand{\herschel}{\mbox{\it Herschel}}
\newcommand{\getsources}{\mbox{\it getsources}}
\newcommand{\simlt}
       {\ifmmode       { \raisebox{-.4em}{$<$}\atop\sim}
          \else        {$\raisebox{-.4em}{$<$}\atop\sim$}
       \fi}

\begin{document} 
\title{The formation of active protoclusters in the Aquila Rift: \\A millimeter continuum view\thanks{Based on observations carried out with the IRAM~30m telescope. 
IRAM is supported by INSU/CNRS (France), MPG (Germany), and IGN (Spain).}}
\author{Ana\"elle Maury\inst{1,2} 
\and Philippe Andr\'e\inst{2} 
\and Alexander Men'shchikov\inst{2}
\and Vera K\"onyves\inst{2}
\and Sylvain Bontemps\inst{3}}
\institute{ESO, Karl Schwarzschild Strasse 2, 85748 Garching bei M\"unchen, Germany
\and Laboratoire AIM, CEA/DSM--CNRS--Universit\'e Paris Diderot, IRFU/SAp, CEA Saclay, 91191 Gif-sur-Yvette, France
\and CNRS/INSU, Laboratoire d'Astrophysique de Bordeaux, UMR5804, BP 89, 33271 Floirac cedex, France } 
\date{Received / Accepted } 
\abstract 
{
While most stars are believed to form in stellar clusters, the formation and early evolution of young stellar clusters is still largely unknown. 
Improving our knowledge of the earliest phases of clustered star formation is crucial for understanding the origin of the stellar initial mass function 
and the efficiency of the star formation process, which both play a key role in the evolution of galaxies.
Here, we present an analysis of the Aquila Rift complex which addresses the questions of the star formation rate (SFR), star formation efficiency (SFE) and typical lifetime of the Class~0 protostellar phase in two nearby cluster-forming clumps: the Serpens South and W40 protoclusters.
We carried out a 1.2~mm dust continuum mapping of the Aquila Rift complex with the MAMBO bolometer array on the IRAM 30m telescope.  
Using a multi-scale source extraction method, we perform a systematic source extraction in our millimeter continuum map.
Based on complementary data from the \herschel\ Gould Belt survey and \spitzer\ maps, we characterize the spectral energy distributions (SEDs)
of the 77 millimeter continuum sources detected with MAMBO and estimate their evolutionary stages.
Taking advantage of the comprehensive dataset available for the Serpens South region, spanning wavelengths from 2~$\mu$m to 1.2~mm, 
we estimate the numbers of young stellar objects (YSOs) at different evolutionary stages and find a ratio of Class~0 to Class~I protostars $N(0)/N(I)=$0.19\,--\,0.27.
This low ratio supports a scenario of relatively fast accretion at the beginning of the protostellar phase, and leads to a Class~0 lifetime of $\sim$~4\,--\,9 $ \times$10$^{4}$~yr. 
We also show that both the Serpens South and W40 protoclusters are characterized by large fractions of protostars  
and high SFRs $\sim$20\,--\,50~$M_{\odot}$Myr$^{-1}$pc$^{-2}$, in agreement with the idea that these two nearby clumps are active sites of clustered star formation 
currently undergoing bursts of star formation, and have the potential ability to form bound star clusters. 
While the formation of these two protoclusters is likely to have been initiated in a very different manner, the resulting protostellar populations are observed to be very similar. 
This suggests that after the onset of gravitational collapse, the detailed manner in which the collapse has been initiated does not affect much the ability of a clump to form stars.
} 
\maketitle 

\section{Introduction}

\noindent{\it{The formation and early evolution of protoclusters}}
\\While most of the young pre-main sequence (PMS) stars in molecular clouds are found in groups and clusters rather than being uniformly distributed throughout the clouds (\citealp{Testi99, Carpenter00, Lada03}, see also \citealt{Bressert10} for a discussion of clustered star formation), 
the formation and early evolution of stellar clusters are still poorly known. 
Improving our knowledge of the earliest phases of clustered star formation holds the key to understanding the distribution of stellar masses at birth since the similarity 
of the stellar initial mass function (IMF) of young stellar clusters with the field stars IMF may imply it is linked 
with the cluster formation process itself \citep{Palla99, Hillenbrand00, Meyer00}. 

Protoclusters are parsec-scale gas-dominated cluster-forming clumps, which contain a large amount of dense gas material (several Jeans masses), and are actively forming protostars, as testified by the significant population of prestellar cores and embedded Class~0 objects detected toward them at (sub-)millimeter wavelengths (see \citealt{Motte98,Hurt96,Myers01,Testi98}). 
While protoclusters are the direct progenitors of infrared embedded clusters, only a small fraction of them eventually result in bound star clusters, as it has been suggested \citep{Lada03} that most embedded clusters are not able to remain bound when they rapidly lose their gas mass and evolve into revealed clusters. Their stellar component therefore disperse to become field stars. 
This high infant mortality has been proposed to be due to insufficient star formation rates (SFR) to fight against fast gas loss \citep{Goodwin09, Baumgardt07}, but may also rely on the dynamical state of the young stars and left-over material during the later evolution of the embedded cluster \citep{SmithR11a}).\\
\indent With ages of only a few $\sim$10$^{5}$~yr in some cases \citep{Peretto07}, protoclusters are therefore the youngest precursors of star clusters. Their youth makes them ideal targets to investigate the formation and early evolution mechanisms of clustered star formation, 
since their properties are likely to still reflect the physics of their formation. 
Deep (sub-)millimeter continuum observations of cluster-forming clumps can set strong constraints on the early properties and conditions 
for clustered star formation \citep[cf.][]{Motte98, Testi98}, the efficiency of the conversion of gas into stars from well-defined mass reservoirs \citep[cf.][]{Enoch07, Belloche11}, 
and the typical lifetimes of the youngest embedded objects, i.e., Class~0 protostars \citep{Andre00, Visser02, Enoch09}. 
Moreover, only (sub-)millimeter line observations can probe the dynamics of cluster-forming clumps \citep{Peretto06, Andre07, KirkH07}, as well as the role of feedback from embedded YSOs on the early evolution of young protoclusters \citep{Knee00, Ginsburg09, Maury09, Nakamura11a}.

Several observational studies targeting both low-mass and intermediate-mass protoclusters have shown that embedded clusters form stars more efficiently than global molecular clouds. Indeed, typical star formation efficiencies (SFEs) of 10--40\% \citep{Wilking83, Olmi02, Lada03} have been estimated for embedded clusters, while molecular clouds have SFEs of only a few percent \citep{Myers86, Evans09, Lada10}. 
The SFE and star formation rate (SFR) are known to vary from region to region, but few studies have focused on comparing these properties for protoclusters formed from different initial conditions.
Moreover, past estimations of SFEs and SFRs have generally relied on only partial population counts and limited source classification, 
as it is very difficult to detect and characterize complete protostellar populations with limited wavelength coverage. 
Indeed, in order to probe complete YSO samples from Class~0 protostars to Class~II objects and 
properly determine their evolutionary stages, deep imaging observations from $\sim$2~$\mu$m to millimeter wavelengths, including the submillimeter domain, 
are required.

The \herschel\ Space Observatory now makes it possible to bridge the submillimeter gap by providing deep, wide-field dust continuum imaging at 
6 wavelengths from 70$\mu$m to 500$\mu$m. 
In particular, the \herschel\ Gould Belt survey \citep{Andre10} is imaging the bulk of the nearest ($d \leq 0.5$~kpc) molecular cloud complexes 
with the SPIRE and PACS cameras and will eventually provide a complete census of Class~0 protostars and prestellar cores in these clouds.

Here, we present a millimeter continuum study of the Aquila Rift complex which addresses the question of the lifetime of the Class~0 phase and 
that of the star formation rate in two nearby cluster-forming clumps: the Serpens South and W40 protoclusters.

~~\\
\noindent{\it{The Aquila Rift complex}}
\\The Aquila Rift molecular cloud complex corresponds to a large extinction feature (see \citealt{Bontemps10}), located above 
the Galactic plane at l=28$^\circ$, only $\sim$3$^\circ$ south of the well-known Serpens Main star-forming region. 
The Aquila Rift extinction-defined area (see \citealp{Bontemps10}) mainly harbours two known sites of star formation: Serpens South is the western 
young embedded cluster revealed by infrared \spitzer\ observations \citep{Gutermuth08}, and W40 is the eastern cluster associated to the eponymous H{\small II} region, also known as Sharpless~2-64 \citep{Smith85,Vallee87}.

A recent Herschel mapping of $\sim$ a $3.3^{\circ} \times 3.3^{\circ}$ field was carried out toward this
complex as part of the Gould Belt survey \citep{Andre10}.
The corresponding far-infrared maps have unprecedented spatial resolution and
sensitivity in 5 bands between 70 and 500 $\mu$m. The first results of these
Herschel observations of the Aquila complex are presented in \citet{Bontemps10}, and \citet{Konyves10}. 
Briefly, the Herschel maps revealed a large number of
new protostars and prestellar cores mainly spread over three currently active star-forming sites: the filamentary molecular cloud harboring the Serpens South infrared embedded cluster in its center, the eastern H{\small II} region W40/Sh2-64, and the southernmost MWC297 / Sh2-62 region. 

~~\\
\noindent{\it{Distance of the star forming clouds in the Aquila Rift}}
\\The distances of the Serpens South and W40 protoclusters, both located inside the extinction feature of the Aquila Rift, are still uncertain. 

Whether or not the Aquila Rift complex and the Serpens Main cloud are part of the same complex and therefore at the same distance is still a matter of debate \citep{Gutermuth08, Dzib10, Loinard11}, however from the extinction maps obtained by \citet{Bontemps10}, their respective extinction features are seen as clearly distinguishable separate regions. 
Indeed, assuming that Serpens Main and Aquila Rift are different entities also has the advantage of explaining the discrepancy found between their extinction-based distance estimates \citep{Straizys96, Straizys03, Bontemps10, Knude11}: $\sim$200--280~pc for the Aquila Rift complex (due to the known extinction wall at a distance of 225$\,\pm\,$55$\,$pc), and $\sim$415~pc for Serpens Main (from the recent VLBI-based distance by \citealt{Dzib10}), suggesting that Serpens Main is located behind the extinction wall associated to Serpens South.

Previous studies targeting the Serpens South protocluster \citep{Gutermuth08, Bontemps10, Konyves10} used the extinction-related distance of $\sim$260~pc for this star forming region, Serpens South being the highest extinction region found inside the Aquila Rift complex.
Also based on extinction measurements \citep{Smith85} and radio spectral line data \citep{Vallee87, Vallee92}, the nearby H{\small II} region W40 has usually been considered to be at a distance ranging from 300 to 700 pc. A recent Chandra study by \citet{Kuhn10} emphasizes, on the basis of X-ray luminosity function arguments, a most-likely distance of $\sim$600~pc  for the W40 cluster, although not excluding the possibility of a smaller ($\sim$300~pc) distance.

We argue that both Serpens South and W40 are likely to be at the same distance because (i) they belong to the same continuous extinction feature (see \citealt{Bontemps10, Schneider11}) associated to the Aquila Rift, and (ii) they are populating a continuum of LSR velocities between 4 and 10~km.s$^{-1}$.
\\For consistency with the \herschel\ Gould Belt survey \citep{Andre10} results for that region \citep{Bontemps10, Konyves10} and easier comparison with previous studies, we adopt a distance of 260~pc for the Aquila Rift complex, including both Serpens South and W40,  throughout this paper.

\section{MAMBO 1.2~mm dust continuum observations and data reduction} 

%%%%%%
\begin{figure*}[!t]
\centering
\includegraphics[width=0.80\textwidth,angle=0,trim=0cm 0cm 0cm 0cm,clip=true]{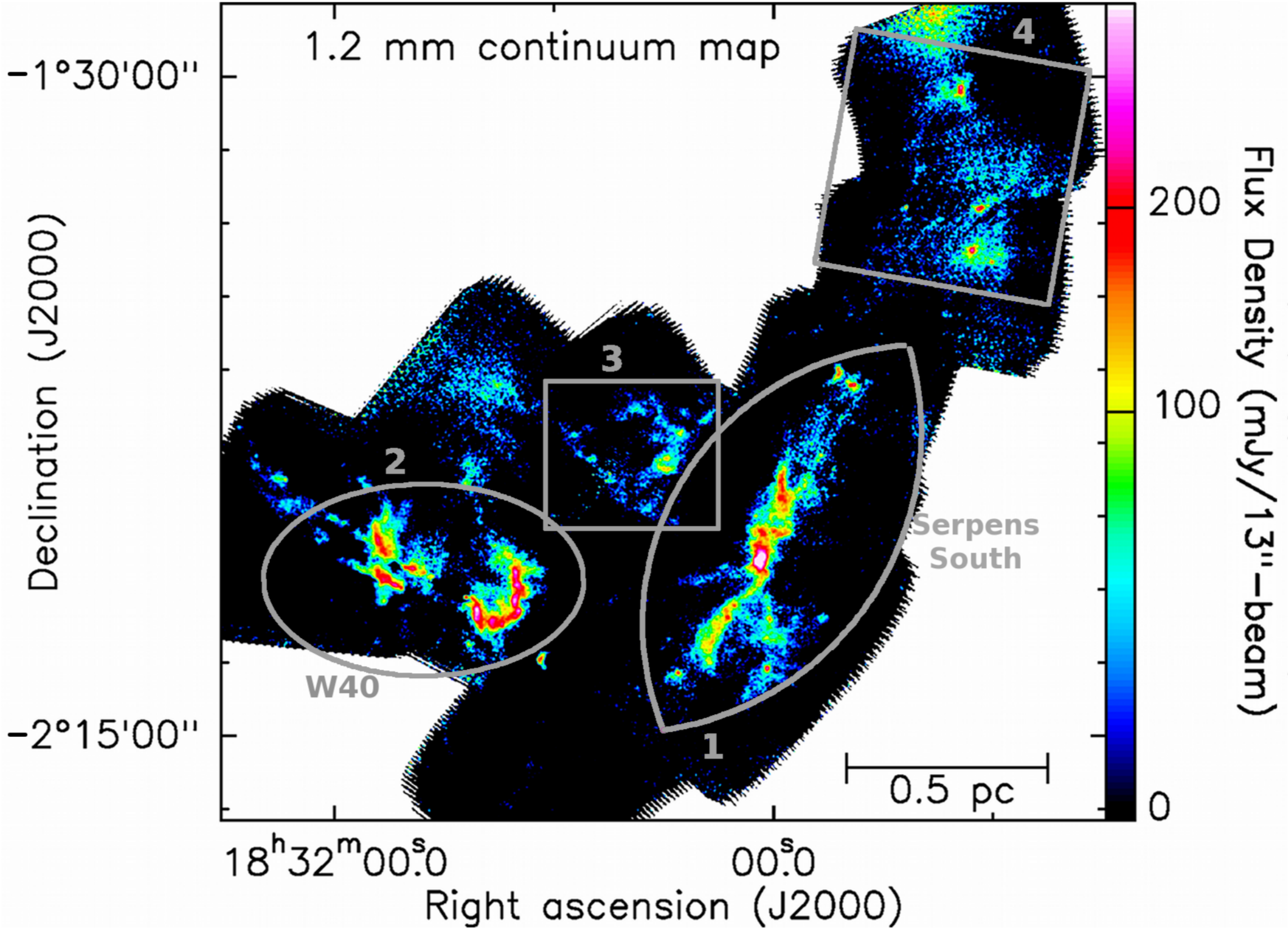}
\caption{1.2 mm dust continuum MAMBO mosaic of the Aquila Rift complex, smoothed to an effective angular resolution of 13\arcsec. Due to the high dynamic range of fluxes in the map, and to enhance visibility of the structure of emission in the whole area mapped, the color scale chosen here has a maximum of 400~mJy/beam. The highest flux density of order $\sim$1400~mJy/beam is found in the center of the Serpens South protocluster (see the blow-up in Figure~\ref{fig:serpenssouth}. In this smoothed map, the mean rms noise level is $\sim$~13~mJy/$13\arcsec$-beam.}
\label{fig:aquila_MAMBO_all}
\end{figure*}
%%%%%%

We used the Max-Planck Millimeter Bolometer array (MAMBO) on the 
IRAM 30-m telescope near Pico Veleta, Spain) during March 2007 
 to carry out a millimeter continuum mapping of the Aquila Rift complex. 
Ten adjacent sub-fields were selected thanks to the near-IR extinction 
map derived from 2MASS data \citep{Bontemps10, Schneider11}, in order to image 
the dust continuum associated with the main extinction features of the Aquila Rift complex, at 1.2~mm ($\sim$240~GHz).
These sub-fields were mapped using the on-the-fly technique in the double-beam switching mode, i.e. 
chopping was performed by switching the wobbling secondary with a throw ranging from 62\arcsec ~to 160\arcsec. 
The scanning speed was 8\arcsec/sec. 
The atmospheric optical depth at zenith was estimated between 0.09 and 0.41 at $\lambda = 1.2$~mm. 
Pointing and focus checks were made every 3 hours, resulting in a pointing accuracy better than 3\arcsec, and the calibration obtained by mapping Mars (primary calibrator) results in a flux accuracy better than $\sim 15\%$. 
The angular resolution with IRAM 30-m antenna is $\sim$ 11\arcsec ~(HPBW) at $\lambda = 1.2$~mm, and the average rms achieved is $\sim$15~mJy/11\arcsec-beam.
We reduced and co-added all the MAMBO sub-fields using standard methods in MOPSIC, part of the GILDAS\footnote{Grenoble Image and Line Data Analysis System, software provided and actively developed by IRAM (\textcolor{blue}{\url{http://www.iram.fr/IRAMFR/GILDAS}})} 
software, and produced a global mosaic covering a total area of $\sim 60\arcmin \times 50\arcmin$ (see Fig.\,\ref{fig:aquila_MAMBO_all}).

\section{Dust continuum mapping: results and analysis} 

\subsection{Main features of the millimeter continuum maps}

The large-scale 1.2~mm mosaic obtained with the MAMBO camera is shown in Fig.\,\ref{fig:aquila_MAMBO_all}. 
Four main parsec-scale regions of interest can be identified in this mosaic, which are labelled {\it{1}}, {\it{2}}, {\it{3}}, and {\it{4}}. 
Indeed, these regions show both extended emission and compact structures in dust continuum, and are therefore likely tracing active star-forming areas. 

\begin{figure}[!h]
\centering
\includegraphics[width=0.97\columnwidth,angle=0,trim=0cm 0cm 0cm 0cm,clip=true]{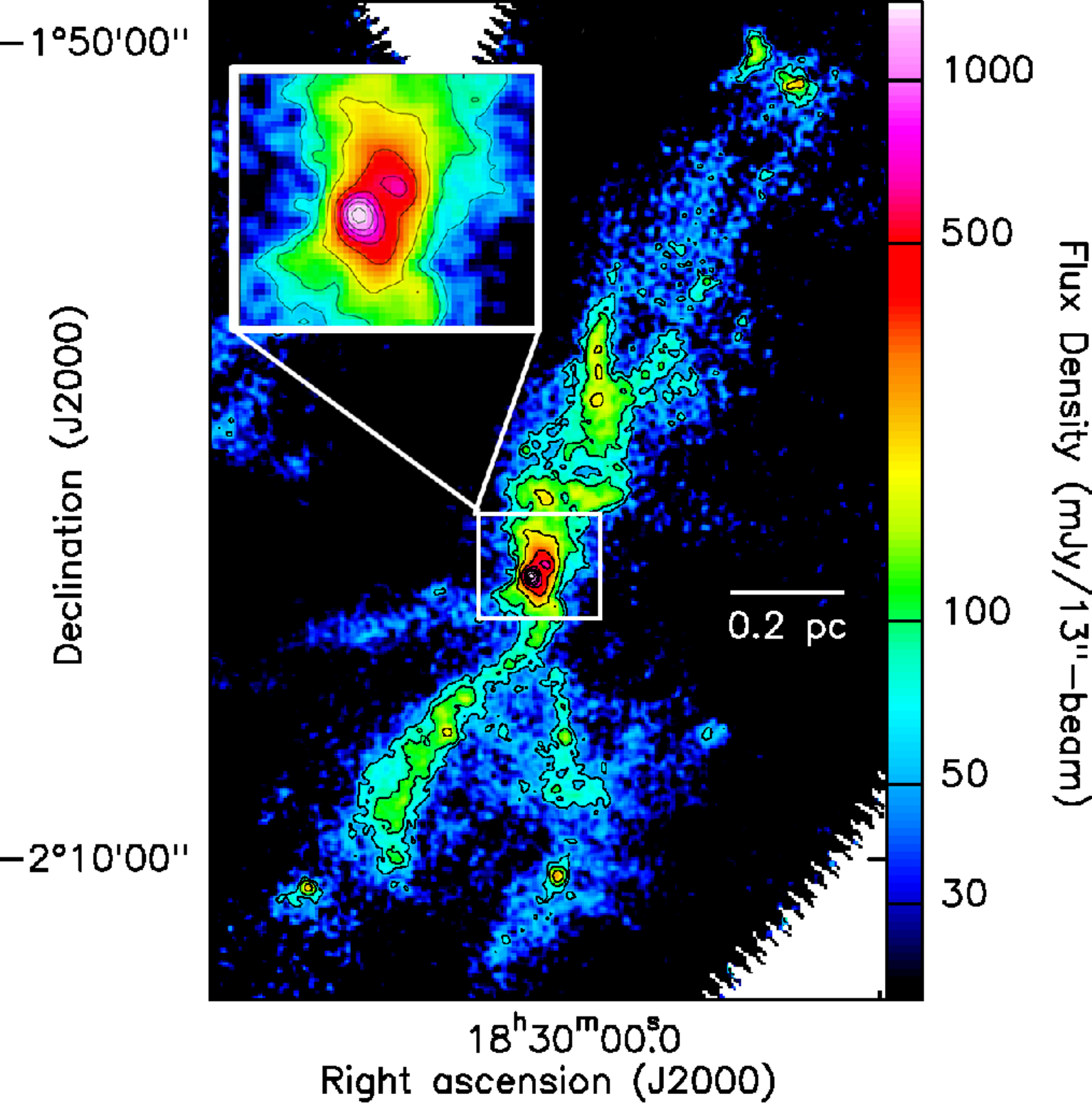}\\
\caption{Blow-up of the Region~1 from the MAMBO 1.2~mm dust continuum map shown in Fig.\,\ref{fig:aquila_MAMBO_all}, around the Serpens South filamentary cloud. The mean rms noise level is $\sim$~12~mJy/$13\arcsec$-beam. Black contours show the levels of 5$\sigma$, 8$\sigma$, 15$\sigma$, 30$\sigma$, 70$\sigma$, 90$\sigma$, and 120$\sigma$ (1.44~Jy/beam).}
\label{fig:serpenssouth}
\end{figure}

Region labelled {\it{1}} is the previously known Serpens South protocluster, showing up as an elongated $\sim$1 pc filament in the 1.2~mm dust continuum map (see blow-up in Fig.\,\ref{fig:serpenssouth}), coincident with the \spitzer\ near-infrared absorption feature discussed by \citet{Gutermuth08}. Very condensed in its center, the filament sees its density decreasing while going along the two arms that have a cross-shape in more diffuse emission. Compact sources are found up to 15\arcmin ~away from the center of the filament.
This high-density region (column densities up to 3$\times$10$^{23}$cm$^{-2}$) at the convergence zone of the filament shows several dust continuum peaks at the MAMBO resolution: it is the center of a protocluster, with several compact millimeter sources tracing very young protostars at the beginning of their accretion phase (separated by less than 5000~AU), as well as numerous more evolved infrared YSOs.

\begin{figure}[!h]
\centering
\includegraphics[width=0.98\columnwidth,angle=0,trim=0cm 0cm 1.3cm 0cm,clip=true]{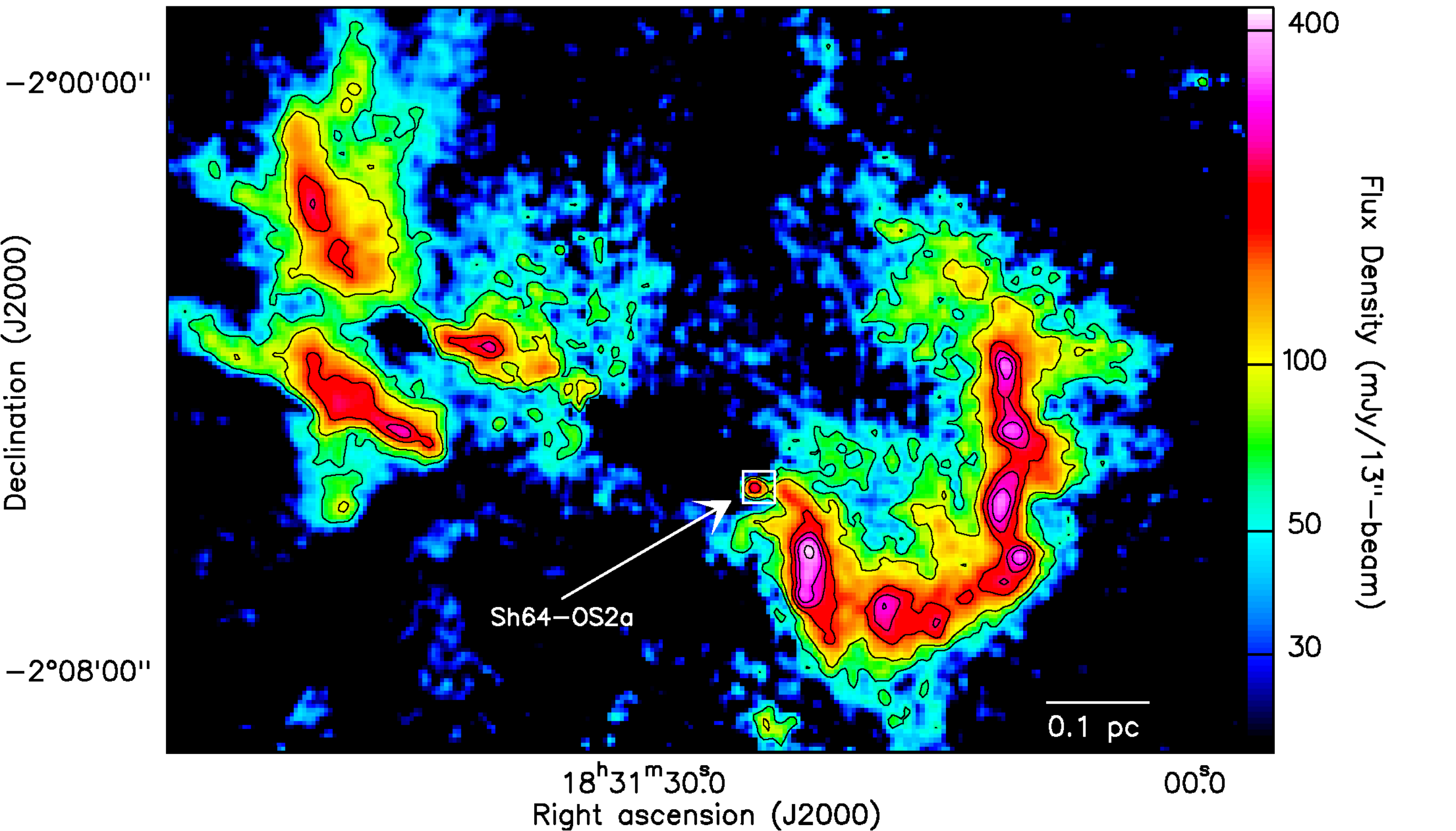}\\
\caption{Blow-up of the Region~2 from the 1.2~mm continuum map shown in Fig.\,\ref{fig:aquila_MAMBO_all}, around the H{\small II} region W40. The mean rms noise level is $\sim$~14~mJy/$13\arcsec$-beam. Black contours show the levels of 5$\sigma$, 8$\sigma$, 15$\sigma$, 20$\sigma$, 25$\sigma$, and 35$\sigma$. One of the candidate exciting stars, Sh64-OS2a, is detected in our MAMBO mapping and indicated here.}
\label{fig:W40_MAMBO}
\end{figure}

Region labelled {\it{2}} is shown with much greater details in Fig.\,\ref{fig:W40_MAMBO}. It mainly features a ring-like structure following the outer border of the W40 H{\small II} region \citep{Smith85,Vallee87,Kuhn10}, and spatially coincident with the CO emission mapped by \citet{Zhu06}. In addition to this western dense shell, the map shows elongated structures (also containing compact sources) in the eastern part of the photo-dissociated region, while the central part of the infrared cluster containing most of the pre-main sequence objects \citep{Kuhn10} does not show any evidence for cold dust emission at 1.2~mm.

Finally, regions labelled {\it{3}} and {\it{4}} are more diffuse, less rich star-formation sites, for which no previous literature was found.

The $\sim 11$\arcsec angular resolution of MAMBO at 1.2~mm allowed us to resolve individual condensations within the dust cores at the assumed distance of Aquila Rift, and we therefore used our millimeter data to complement the \herschel\ dataset and characterize the protostellar population in the mapped area. 

%%%%%%
%%%%%%

%%%%%%
%%%%%%

\subsection{Source extraction and properties}

In order to improve the detection of compact sources in our field, we smoothed the original 1.2~mm continuum map to a 13\arcsec-beam, and performed a systematic source detection on the MAMBO map using the new source extraction method \getsources\ 
\footnote{This new algorithm analyzes fine spatial decompositions of original images across a wide range of scales and across all 
wavelengths, cleans those decomposed images of noise and background fluctuations separately at each wavelength, and
constructs a set of wavelength-independent detection images that preserve information in both spatial and wavelength 
dimensions as well as possible. Sources are detected in the set of decomposed combined detection images by following 
the evolution of their segmentation masks across all spatial scales. Measurements of the sources' properties are 
performed in the original images at each wavelength after the background has been subtracted by interpolation under 
the sources "footprints" and overlapping sources have been deblended, see Men'shchikov et al. 2011 (in prep).
} \citep{Menshchikov10,Menshchikov11}. 
This analysis led to the extraction of a total of 77 dust continuum sources in the whole region shown in Fig.\,\ref{fig:aquila_MAMBO_all}, among which 25 are located in the Serpens South filament, while 36 lie in the W40 region. 
The resulting MAMBO compact sources catalog is reported in Table~1. 
Starting from this 1.2~mm dust continuum set of sources, parallel measurements were performed at other wavelengths with \getsources\ in order to estimate the counterparts 
of the MAMBO sources in both the Herschel (from 70$\mu$m to 500$\mu$m), the Spitzer (8$\mu$m and 24$\mu$m) and the 2MASS K-band maps, when data was available. All these sources were also examined individually by eye, ensuring association to the right counterparts, which is not always straightforward while working with different sizes of beams in crowded regions, such as Serpens South or W40.

We used these flux measurements to establish spectral energy distributions (SEDs) for all the 77 MAMBO sources, and derived other source properties (dust temperature, core/envelope mass, bolometric temperature and luminosity) when a good fit was found to reproduce the SED obtained between 2$\mu$m and 1.2~mm with grey-body functions.
The condensation masses (core masses for starless sources and envelope masses -- $M_{\rm{env}}$ -- for protostellar objects) 
and dust temperatures T$_\mathrm{dust}$ were estimated from the fitting of these SEDs by grey-body functions \citep{Konyves10}. We adopted the recommended values \citep{Ossenkopf94, Henning95} for the dust opacity per unit mass column density (of dust and gas): $\kappa_{\rm{1.2mm}} = 0.01~{\rm cm}^{2} \, {\rm g}^{-1}$ for protostellar envelopes, and $\kappa_{\rm{1.2mm}} = 0.005~{\rm cm}^{2} \, {\rm g}^{-1}$ for starless clumps.
The condensations have masses ranging from $\sim 0.02~M_\odot$ to $\sim 6~M_\odot$, with an average factor $\sim 2$ absolute uncertainty (see Table~1). Note also that for the most evolved sources (mainly Class~I protostars), these masses might be contaminated or even dominated by disk material, while they represent the core masses for the starless sources.

The MAMBO sources showing no counterpart shortward of 160~$\mu$m with \herschel\--PACS, but detected in at least two \herschel\--SPIRE bands (250, 350, or 500$\mu$m) were classified as starless sources. 
The MAMBO sources with compact counterparts at \herschel\ wavelengths $< 160~\mu$m were classified as candidate young stellar objects (YSOs), 
The bolometric luminosities $L_{\rm{bol}}$, submillimeter luminosities $L_{\rm{smm}}^{\lambda > 350\mu}$, and bolometric temperatures $T_{\rm{bol}}$ of the candidate YSOs  
were estimated by integrating their SEDs, and are reported in Table~1. 
Note that the values of $L_{\rm{smm}}^{\lambda > 350\mu}$ and $L_{\rm{bol}}$ derived here are expected to be more accurate (estimated uncertainty $\leq$40\%) 
than previous estimates from the literature, thanks to the \herschel\ data providing a good coverage of the far-infrared/submillimeter domain.

\subsection{Evolutionary classification of protostellar sources}

%%%%%%
\begin{figure*}[!t]
\centering
\includegraphics[width=0.60\textwidth,angle=0,trim=0cm 0cm 0cm 0cm,clip=true]{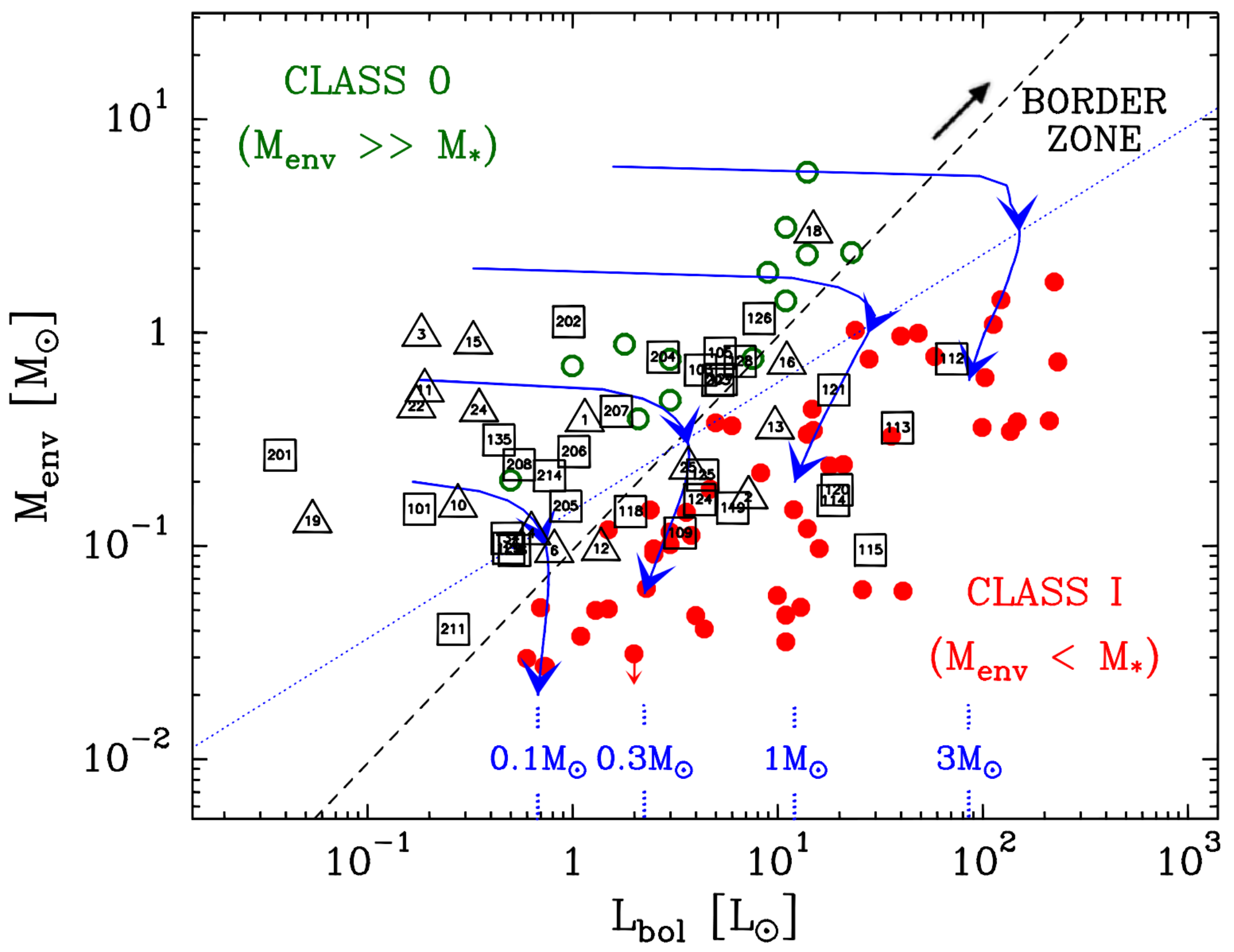}
\caption{Envelope mass versus bolometric luminosity diagram comparing the locations of the protostars detected with MAMBO in the Aquila Rift region 
(open triangles for the Serpens South sub-field, open squares for the other sub-fields) with the positions of knwon low-mass Class~I (filled circles) and 
Class~0 objects (open circles) in Ophiuchus, Perseus, and Orion \citep{Andre00}. 
Numbers reported inside triangles are the reference numbers of the sources reported in the last column of Table~1.
Model protostellar evolutionary tracks, computed for various final stellar masses (indicated above the $L_{\rm{bol}}$ axis) 
assuming the accretion/luminosity history described in the text, are superimposed (4 blue tracks representing protostellar evolution 
for objects with final masses 0.1, 0.3, 1 and 3M$_{\odot})$. 
Evolution proceeds from the upper left to the lower right as indicated by two arrows on each track,
plotted when 50\% and 90\% of the final stellar mass has been accreted, respectively.
The straight  lines show two $M_{\rm{env}}$--$L_{\rm{bol}}$ relations marking the conceptual border zone 
between the Class~0 ($M_{\rm{env}} > M_\star  /\epsilon $) and the Class~I ($M_{\rm{env}} < M_\star /\epsilon $) stage, 
where $\epsilon $ is the local star formation efficiency taken here to be 50\%. 
The dashed line is such that $M_{env} \propto L_{\rm{bol}}$ \citep[cf.][]{Andre93,Andre94a}, 
while the dotted relation follows $M_{\rm{env}} \propto L_{\rm{bol}}^{0.6}$ as suggested by 
the accretion scenario adopted in the tracks (see also \citealp{Bontemps96a}). 
The typical uncertainty in the plotted values of $L_{\rm{bol}}$ is $\sim$ 30 -- 40\%, 
while the uncertainty in the $M_{\rm{env}}$ values is a factor of $\sim 2$ (see \S ~3.2). 
The black arrow in the upper right indicates how the positions of the W40 protostars would move in the diagram if a distance of 600 pc was adopted for W40, instead of 260 pc.
}
\label{fig:Menv-Lbol_diagram}
\end{figure*}
%%%%%%

All of the candidate embedded protostellar objects detected with MAMBO were then placed in the envelope mass versus bolometric luminosity 
($M_{\rm{env}}$--$L_{\rm{bol}}$) diagram shown in Fig.\,\ref{fig:Menv-Lbol_diagram}, in order to distinguish between envelope-dominated 
($M_{\rm{env}} >> M_{\star}$) Class~0 protostars and more evolved (Class~I) protostars. 
In this diagram, the bolometric luminosity of a protostellar object can be used as an indirect tracer of the stellar mass 
$M_{\star}$ it has already accreted \citep{Andre93, Andre94a}, based on comparison with simple accretion model tracks 
(cf. \citealt{Bontemps96a, Saraceno96, Andre00}).
\\The specific evolutionary tracks shown in Fig.\,\ref{fig:Menv-Lbol_diagram} are similar to the tracks used by \citet{Andre08b} 
and were computed assuming each protostar forms from a prestellar core of finite initial mass $M_{\rm{core}} \equiv M_{\rm{env}}(0) $
and has $L_{\rm{bol}} = GM_{\star}(t)\dot{M}_{\rm{acc}}(t)/R_{\star}(t) + L_{\star}(t)$, where 
$R_{\star}(M_{\star})$ is the protostellar radius and $L_{\star}$ the PMS birthline luminosity \citep{Stahler88}.
The mass accretion rate and the envelope mass were assumed to be related by $\dot{M}_{\rm{acc}} (t) = \epsilon\,M_{\rm{env}}(t)/\tau$, 
where $\epsilon = 50\% $ is the typical star formation efficiency for individual cores (cf. \citealt{Matzner00})
and $\tau = 10^5$~yr is the characteristic timescale of protostellar evolution, leading to $\dot{M}_{\rm{acc}} (t)$ and 
$M_{\rm{env}}(t) $ functions declining exponentially with time as suggested by the observed decline of outflow power from Class~0 to Class~I (see \citealp{Bontemps96a}). 
Note that we adopt $\epsilon = 50\% $ here as in \citet{Andre08b} instead of $\epsilon = 100\% $ as in 
\citet{Andre00}, since recent comparisons of the prestellar core mass function with the stellar IMF suggest 
$\epsilon \simlt 50\% $ (e.g. \citealt{Alves07, Nutter07, Konyves10, Andre10}). 

A Class~0 status was attributed to sources with $M_{\rm{env}} > M_{\star}/ \epsilon$ according to the diagram of Fig.\,\ref{fig:Menv-Lbol_diagram}, 
i.e., protostars which have accreted less than half of their final stellar mass. 
Conceptually, protostars at the border line between the Class~0 and the Class~I stage are objects which have accreted exactly half of their final stellar mass, 
and which thus have $M_{\rm{env}} = M_{\star}/ \epsilon$ according to the tracks.
In practice, sources falling into the  "Border Zone" region located between the dashed and dotted lines in Fig.\,\ref{fig:Menv-Lbol_diagram} 
have been classified as border-line Class~0/I objects in Table~1.
Sources lying below this border zone in the diagram were classified as Class~I or Class~II, depending on their infrared spectral 
index $\alpha_{\rm{IR}} \equiv d{\rm log}\lambda F_\lambda/d{\rm log}\lambda$ 
between 2.2$\mu$m and 24$\mu$m (as estimated by us from independent \getsources\ measurements 
at 2.2$\mu$m, 8$\mu$m, 24$\mu$m when available -- see \S4).
\\The use of this evolutionary diagram (see Fig.\,\ref{fig:Menv-Lbol_diagram}) 
leads to the identification of 9 Class~0 (and 3 border-line Class~0/I) protostars 
in the Serpens South filament. Likewise, 8 Class~0 (and 4 border-line Class~0/I) protostars are found in the W40 region. 
The other sub-fields labelled {\it{3}} and {\it{4}} gather 8 Class~0 (and 2 border-line Class~0/I) protostars.

%%%%%%
\begin{figure*}[!t]
\centering
\includegraphics[width=0.60\textwidth,angle=0,trim=0cm 0cm 0cm 0cm,clip=true]{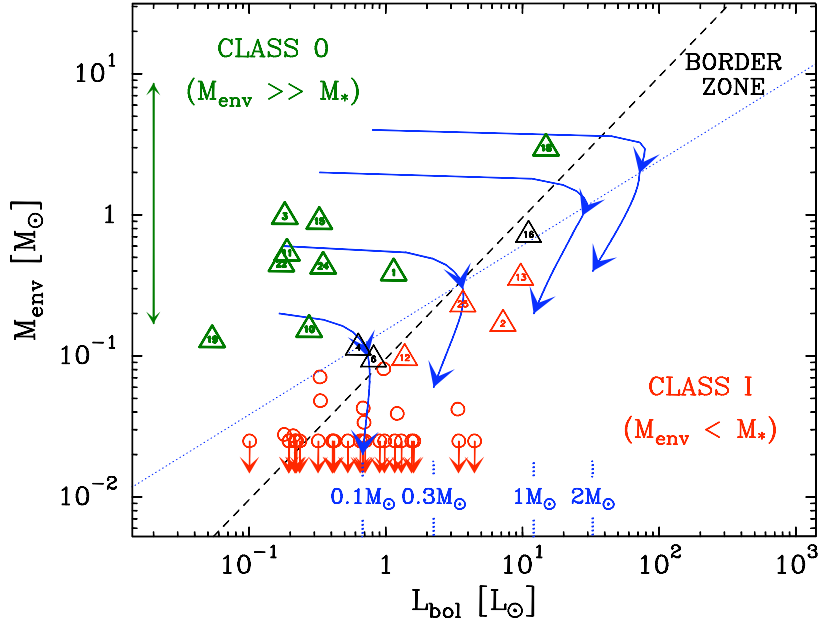}
\caption{
Same as Figure~4, but showing only the sources from the Serpens South cluster forming clump. Triangles show the locations of the MAMBO protostars (green triangles are Class~0 and red triangles are Class~I protostars, while black triangles represent the border-line objects). Red circles show the locations of the Class~I protostars detected by \spitzer~ and without MAMBO detection. The Class~I sources without any MAMBO counterpart are indicated as upper-limit values for M$_{\rm{env}}$, around the detection limit of 2.5$\times$10$^{-2} M_\odot$. The green double arrow along the $M_{\rm{env}}$ axis shows the range of masses for the prestellar cores detected by \herschel, based on a preliminary analysis of the Gould Belt survey data, which are 85\% complete down to  0.3$M_\odot$ in this region \citep{Konyves10}. The respective locations of the prestellar cores, Class~0 and Class~I protostars in the Serpens South protocluster are roughly consistent with what is expected from a single population of objects evolving along the model evolutionary tracks shown in blue.
}
\label{fig:Menv-Lbol_zoom}
\end{figure*}
%%%%%%

Two other criteria, [$L_{\rm{submm}}^{\lambda > 350\mu}/L_{\rm{bol}} > 1\%$]\footnote{The original criterion introduced by \citet{Andre93} was 
$L_{\rm{submm}}^{\lambda > 350\mu}/L_{\rm{bol}} > 0.5\%$, but only YSOs with $L_{\rm{submm}}^{\lambda > 350\mu}/L_{\rm{bol}} > 1\%$ 
were considered bona-fide Class~0 objects by \citet{Andre00}, while sources with $L_{\rm{submm}}^{\lambda > 350\mu}/L_{\rm{bol}} $
between $\sim 0.5$\% and $\sim 1$\% were classified as border-line objects \citep[see Table 1 of][]{Andre00}.} 
\citep[e.g.][]{Andre00} and  [$T_{\rm{bol}} < 70$~K] \citep[e.g.][]{Myers98a} respectively, 
have been commonly used in the literature to select candidate Class~0 protostars among embedded YSOs.
In order to assess the importance of uncertainties in our classification, we have applied these two alternative methods of classifying embedded YSOs to the Aquila Rift sample 
and compared the results to those obtained above on the basis of the $M_{\rm{env}}$ -- $L_{\rm{bol}}$ diagram. 
\\We find (see Table~1) that using the $L_{\rm{submm}}^{\lambda > 350\mu}/L_{\rm{bol}} > 1\%$ criterion 
(which traces a ratio of envelope to stellar mass larger than unity -- see \citealt{Andre93}) 
to select Class~0 protostars leads to similar results to those obtained from the $M_{\rm{env}}$ -- $L_{\rm{bol}}$ diagram. 
Only in a few cases does the use of this criterion lead to the selection of border-line Class~0/I sources (according to $M_{\rm{env}}$ -- $L_{\rm{bol}}$ diagram) 
as candidate Class~0 protostars. 
For example, using the $L_{\rm{submm}}^{\lambda > 350\mu}/L_{\rm{bol}} > 1\%$ criterion leads to the selection of 12 Class~0 protostars in the Serpens South filament 
(compared to 9 Class~0 and 3 border-line sources when using the $M_{\rm{env}}$ -- $L_{\rm{bol}}$ diagram), 
and 11 Class~0 protostars in the W40 region (compared to 8 Class~0 and 4 border-line sources when using the $M_{\rm{env}}$ -- $L_{\rm{bol}}$ diagram).
Based on the present study, we suggest that setting the border line at $L_{\rm{submm}}^{\lambda > 350\mu}/L_{\rm{bol}} > 3\%$ may better match the results obtained with the more conceptual 
classification based on the tracks in the $M_{\rm{env}}$ -- $L_{\rm{bol}}$ diagram that we use here.
\\We find that using the $T_{\rm{bol}} < 70$~K criterion for distinguishing Class~0 protostars from Class~I objects \citep{Chen95, Myers98a} tends to systematically overestimate the number of candidate Class~0 objects, compared to the numbers obtained from both the $M_{\rm{env}}$ -- $L_{\rm{bol}}$ diagram and the $L_{\rm{submm}}^{\lambda > 350\mu}/L_{\rm{bol}} > 1\%$ criterion. 
This is especially true in the "warm" W40 region where the estimates of $T_{\rm{bol}}$ are more uncertain due to confusion at mid-infrared wavelengths from the photo-dissociation region. 
For example, using the $T_{\rm{bol}} < 70$~K criterion leads to the identification of 12 Class~0 protostars in the Serpens South filament, and 19 Class~0 protostars in the W40 protocluster (twice the number of Class~0 protostars found in this region if using the two other criteria presented above).
\\We therefore conclude, in agreement with \citet{Young05}, that the observational criterion to distinguish Class~0 protostars making use of the 
$L_{\rm{submm}}^{\lambda > 350\mu}/L_{\rm{bol}}$ ratio seems to be more reliable than the $T_{\rm{bol}}$ criterion, also because it is less sensitive to environment and geometry effects. 
Comparing the various methods of classifying protostars, which is roughly equivalent to taking into account the border-line Class~0/I protostars from the $M_{\rm{env}}$ -- $L_{\rm{bol}}$ diagram, 
we find that the typical uncertainty in the number of Class~0 protostars is  $\sim$50\% in W40, and $\sim$30\% in Serpens South.

\section{Discussion and conclusions}

\subsection{Class~0 phase lifetime and star formation rates}

Here we propose a statistical analysis of the properties of the protostellar populations in the two main regions showing active star formation from our MAMBO map: the Serpens South filamentary cloud and the W40 H{\small II} region.

\subsubsection{The Serpens South protocluster}

\indent First, we take advantage of having a complete population of Class~0 protostars in a spatially well-defined star-forming cloud to derive an estimate of the lifetime of the Class~0 phase.

The Serpens South protocluster was observed with IRAC and MIPS onboard the \spitzer\ telescope, and the analysis of IRAC data by \citet{Gutermuth08} revealed 37 Class~II YSOs and 54 Class~I protostars (including flat spectrum sources) in a region of 14\arcmin$\times$10\arcmin ~around the filament. 
Based on the publicly available MIPS-24$\mu$m, IRAC-8$\mu$m, and 2MASS/$K$-band data,  
we independently performed an extraction of MIPS-24$\mu$m point-like sources with \getsources\ in the $0.8 \times 1.2$~pc
area around the Serpens South filament shown in Fig.\,\ref{fig:serpenssouth}, which revealed a total of 112 objects  with an infrared spectral index 
$\alpha_{\rm{IR}} \equiv d{\rm log}\lambda F_\lambda/d{\rm log}\lambda > 0.3$ between 2$\mu$m and 24$\mu$m, that we considered as candidate Class~I protostars.
\\As discussed in \S ~3.2 above, 
our MAMBO mapping finds 9 candidate Class~0 protostars in the same area (not counting border-line Class~0/Class~I sources). 
Our census of Class~0 protostars in Serpens South is sensitive to protostars with final masses down to $\sim 0.1$~M$_{\odot}$ (see evolutionary tracks in Fig. ~\ref{fig:Menv-Lbol_diagram}).
According to the grid of protostellar models by \citet{Robitaille07}, Class~I protostars with final masses $M_{\star}^{\rm{final}} >$ 0.1~$M_{\odot}$ are expected to 
have $24\mu$m flux densities $F_{24\mu \rm{m}} > 35$~mJy at $d = 260$~pc (assuming a random distribution of inclination angles, i.e computed using a mean value of $\sim$57$^{\circ}$). 
We will therefore consider here only the 48$\pm$2 candidate Class~I sources which have $F_{24\mu \rm{m}} > 30-40$~mJy from our extracted sample. 
Note that at such a $24\mu$m flux density level, our Class~I sample should not be contaminated by extragalactic sources \citep[cf.][]{Harvey07}.
Among these 48 objects, 16 coincide with 21 IRAC Class~I sources reported by \citet{Gutermuth08}, 
while 21 are new detections made possible thanks to the use of the 24~$\mu$m data.
The \spitzer, \herschel, and MAMBO maps are very likely wide enough to probe complete samples of YSOs associated with the Serpens South 
filamentary clump, since there are no YSOs, neither high column-density gas, immediately outside this $0.8 \times 1.2$~pc area (shown in Fig.\,\ref{fig:serpenssouth}) in 
available larger \herschel, \spitzer, and MAMBO maps of the region. 
\\Assuming that the star formation rate in Serpens-South has been constant over the duration of the protostellar phase, 
we can estimate the lifetime of the Class~0 phase by comparing the relative numbers of Class~0 and Class~I sources in this region.
Combining the MAMBO and \spitzer\ samples discussed here, 
the Serpens South  cluster-forming filament harbors a total of 57 protostellar objects with expected final stellar masses $>$ 0.1~$M_{\odot}$, and more precisely 84\% of Class~I and 16\% of Class~0 objects.
Adopting a lifetime of 0.2--0.5~Myr for the Class~I phase \citep{Greene94, Evans09}, the ratio of Class~0 to Class~I protostars we find, $N(0)/N(I)=0.19$,  
suggests a typical lifetime of $\sim$~4--9 $\times10^{4}$~yr for the Class~0 phase.
This result is intermediate between the early estimate of 1--3$\times$10$^{4}$~yr by \citet{Andre00} and the more recent estimate of 0.10--0.16~Myr by \citet{Evans09}. 

Our analysis makes use of an unprecedented census of YSOs in Serpens South,  
using wavelengths from 2~$\mu$m to 1.2~mm and including the \herschel\ submillimeter domain. 
It should therefore provide a more complete population of embedded YSOs in Serpens-South and 
a more reliable assessment of their evolutionary states (via better determination of their envelope masses and bolometric luminosities) 
than earlier studies. 
\\However, the lifetime we derive for the Class~0 phase is still subject to large uncertainties, partly due to remaining uncertainties on the lifetime of the Class~I phase.  
Estimates of the latter are themselves highly dependent on assumptions  concerning the lifetime of the Class~II (or T Tauri) phase (see \citealt{Evans09}). 
\\The method chosen to distinguish Class~0 protostars from more evolved YSOs also impacts the determination of this lifetime, because different methods lead to slightly different source counts (see \S~3.3). However, the two favored methods to classify YSOs were compared in \S~3.3, and the differences they show (typical $\sim$30\% uncertainty on the number of Class~0 protostars detected, i.e $N(0)/N(I) \sim 0.19-0.27$) are expected to result in a typical uncertainty of $\sim$0.02~Myr in the Class~0 lifetime we find here, which is a factor of 2 lower than the main source of error due to the uncertainty in the Class~I lifetime (leading to a $\sim$0.05~Myr uncertainty on the Class~0 lifetime as shown above).
Moreover, we do not claim our result to be universal, as our study is focused on only one star-forming region and does not address the question of environmental effects.
\\Even taking these caveats into account, we stress that our results tend to support a scenario of relatively fast accretion at the beginning of the protostellar phase 
in protoclusters (cf. \citealt{Henriksen97}), since they suggest that a typical YSO accretes the first half of its final stellar mass in $<$ 0.1~Myr (Class~0 phase) 
and the second half in $\sim$ 0.4~Myr (Class~I phase -- cf. \citealt{Evans09}).
\\Finally, we remark, in Fig.~\ref{fig:Menv-Lbol_zoom}, an over-abundance of low-luminosity Class~I protostars if compared to what is expected from the population of Class~0 protostars and associated model evolutionary tracks in the $M_{\rm{env}}$--$L_{\rm{bol}}$ diagram for Serpens South, which is likely to be an illustration of the well-known "luminosity problem" first noted by \citet{Kenyon90} and more recently stressed by \citet{Evans09}. However, we point out that the locations of prestellar cores, Class~0 and Class~I protostars in the $M_{\rm{env}}$--$L_{\rm{bol}}$ diagram for Serpens South shown in Fig.\,5, are roughly consistent with what expected for a single population of objects evolving along the model evolutionary tracks. This illustrates the validity of our method for determining the evolutionary status of protostars.

Second, using the column-density map of the Aquila Rift derived from \herschel\ data \citep{Konyves10}, we can estimate a mass for the Serpens South filamentary cluster-forming clump 
shown in Fig.\,\ref{fig:serpenssouth}, and derive a value of the star formation efficiency (SFE) for this region. 
We find a total mass of $\sim$610 $M_{\odot}$ in the 15\hbox{$^{\prime}$} $\times$ 25\hbox{$^{\prime}$} area containing the protostellar population discussed here, with a mean column density of $N_{H_2} \sim 3.1\times10^{22}\,  {\rm cm}^{-2}$. 
Considering that this region has formed 9 Class~0,  48 Class~I, and 37 Class~II objects, i.e. a total of 94 YSOs of mean mass 0.5~$M_{\odot}$, we find:
SFE=$M_{star}/(M_{star}+M_{gas})=47/(47+610) \sim$ 7\%.
Assuming a standard duration of 2~Myr for going from the Class~0 to the Class~II phase \citep{Evans09} and a projected area of $\sim$1~pc$^{2}$ for the region harboring YSOs, we derive the following star formation rate per unit surface for the Serpens South cluster-forming clump: SFR$\sim$23 $M_{\odot}$ Myr$^{-1}$pc$^{-2}$. 
\\This value is significantly higher than the typical SFR values reported by \citet{Evans09} for entire molecular clouds (SFR$\sim$1.6 $M_{\odot}$ Myr$^{-1}$pc$^{-2}$), 
which is consistent with the fact that we are focusing here on a compact cluster-forming clump with {higher column densities (up to $N_{H_2} \sim 10^{23}\,  {\rm cm}^{-2} $) than the clouds targeted by \citet{Evans09}, 
also showing evidence of a recent burst of star formation. 
A recent burst of star formation is indicated by the unusually large number of Class~0 and Class~I protostars found in the Serpens South star-forming clump 
compared to the relatively small number of Class~II YSOs, which was already pointed out by \citet{Gutermuth08} and that we confirm here. 
Note that if Serpens-South has indeed undergone a recent burst of star formation, then the observed ratio of 
Class~0 to Class~I protostars ($N(0)/N(I)$) may overestimate the actual ratio of Class~0 to Class~I lifetimes.
\\Finally, the SFE value derived here for Serpens South is of the same order as that reported by \citet{Evans09} for molecular clouds (i.e. SFE$\sim$3--6\%). 
The fact that Serpens South shows both a high SFR and a relatively low SFE may be interpreted as an additional sign of its very early evolutionary stage \citep{Lada03}.

\subsubsection{The W40 H{\small II} region}

The present MAMBO 1.2~mm continuum study allows us to identify 8 candidate 
Class~0 protostars in the W40 cluster region, where copious photo-dissociation processes (H{\small II} region probably powered by the massive young star Sh64-OS2a, see Fig.~\ref{fig:W40_MAMBO}) make it difficult to obtain a census of YSOs at shorter wavelengths. Indeed, the Lyman-continuum luminosity towards the H{\small II} region of W40 ($\sim$1.5$\times$10$^{48}$ photons.s$^{-1}$) inferred from the radio continuum emission by \citet{Altenhoff70} could be produced by a single O9 V star \citep{Smith85}.
In a recent study, \citet{Kuhn10} found 93 YSOs with NIR $K$-band excess belonging to the W40 region, based on 
their $Chandra$ observations toward the young cluster. 
However, their census did not distinguish between Class~I, Class~II, and later evolutionary stages. 
It is therefore difficult to derive accurate lifetimes for the Class~0 and/or Class~I phases in the W40 region.

We estimate a gas mass of 310~$M_{\odot}$ from the \herschel\ column density map \citep{Konyves10} of the region shown in Fig.~\ref{fig:W40_MAMBO}, similar to what was found (320~M$_{\odot}$) from the $^{13}$CO(2-1) map by \citet{Zhu06}.  
The mean column density over this region is $N_{H_2} \sim 1.8\times10^{22}\,  {\rm cm}^{-2}$.
Considering that this cluster presently contains a total of 20 protostellar sources detected with MAMBO and 93 Class~II sources 
showing NIR $K$-band excess and detected with $Chandra$ \citep{Kuhn10}, 
%i.e., a total of 113 YSOs, 
we estimate that the current star formation efficiency of the protocluster is SFE = 56/(310+56) $\sim$ 15\% assuming a mean stellar mass of 0.5~$M_{\odot}$ 
\footnote[4]{Note that if we take into account the additional Class~III objects identified in X-rays by \citet{Kuhn10} ($\sim$109 $Chandra$ YSOs without $K$-band excess ), then the SFE goes up to $\sim$26\% and the upper-limit of the SFR to $\sim$120 M$_{\odot}$ Myr$^{-1}$pc$^{-2}$ (assuming a lower-limit age of $\sim$1~Myr, see \citealt{Kuhn10}).}.
We stress here that the mass derived from the \herschel\  submillimeter continuum maps may be significantly lower than 
the original clump mass reservoir that formed the PMS objects presently observed, as a fairly large fraction of the gas may have 
already been expelled from the region. This is suggested for example by the lack of any submillimeter/millimeter continuum emission toward the center 
of the cluster where most of the PMS stars are detected.

Assuming a distance of 260~pc for W40, the projected area of the W40 region harboring YSOs (area shown in Fig.~\ref{fig:W40_MAMBO}) is 0.9 pc$^{2}$. 
Assuming a period for star formation of 2~Myr (up to the Class~II population), we can derive a lower-limit value for the SFR of 34 M$_{\odot}$ Myr$^{-1}$pc$^{-2}$, 
which is similar to what is found for the Serpens South protocluster (see \S4.1.1).
However, \citet{Kuhn10} argue that the high disk fraction they find thanks to $K$-band excess measurements would support an age for the W40 cluster of only $\sim$1~Myr, 
which would lead to a upper-limit SFR\footnotemark[4] of $\sim$57~$M_{\odot}$Myr$^{-1}$pc$^{-2}$. 
For comparison, the star formation rate in the Serpens Main cluster, i.e. 56~$M_{\odot}$ Myr$^{-1}$ pc$^{-2}$ \citep{Harvey07}, is of the same order of magnitude as the values we find here. 

We note that the classification we adopted here to determine the evolutionary status of the MAMBO sources, based on the $M_{\rm{env}}$ -- $L_{\rm{bol}}$ diagram shown in Fig. ~\ref{fig:Menv-Lbol_diagram}, is independent of the distance used for W40 as the envelope mass and bolometric luminosity have the same dependency with distance.
At first sight, we may expect the SFR and SFE values
estimated for the W40 complex to scale as $1/d^{2}$, since both the gas mass and
the cloud area scale as $d^{2}$, while the number of YSOs detected
stays the same. However, assuming a standard IMF, the total stellar mass
is roughly proportional to the total bolometric luminosity of the YSO
population, which also scales as $d^{2}$. Therefore, we argue that both
the SFR and the SFE we estimate for W40 do not depend much on the
adopted distance.

\subsection{Comparing evolutionary models and IMF predictions in our Serpens South sample}

Considering that we detected 57 protostellar objects with final masses $M_{\star}>$ 0.1 $M_{\odot}$ in the Serpens South cluster-forming clump, 
and following the IMF \citep{Kroupa01,Chabrier03a}, one would expect the most massive object from our sample to have $M_{\star,max}^{\rm{final}} \sim$ 2.5$\pm$0.5 M$_{\odot}$ (taking into account the $\sim$30\% uncertainty on the number of protostellar objects in Serpens South). 
Similarly, applying the IMF to our sample of YSOs predicts a median final mass $\overline{M_{\star}^{\rm{final}}} \sim$ 0.35$\pm$0.1 $M_{\odot}$, for our sample. 
\\The masses of most of the Class~I sources are not constrained by our MAMBO observations (43 out of 48 are \spitzer\ sources not detected with MAMBO). Moreover, variable accretion rates during the embedded phases (only scenario likely to solve the luminosity problem, see \S3.3, seen in our data by an over-abundance of low-luminosity Class~I protostars in Fig.~\ref{fig:Menv-Lbol_zoom}), do not allow us to use luminosities to derive accurate masses for Class~I protostars not being detected at sub-millimeter or millimeter wavelengths. Therefore, we use the properties from our sample of Class~0 protostars to compute the median stellar mass and maximum stellar mass, expected from the model evolutionary tracks of Fig.\,\ref{fig:Menv-Lbol_zoom}. We then compare these values with values expected from the standard IMF. We stress however that the low number of Class~0 protostars we use here is not allowing to draw robust conclusions.
\\We find a median envelope mass $\overline{M_{\rm{env}}}\sim$ 0.8\,$M_{\odot}$. If we follow the evolutionary model tracks shown in Fig.\,\ref{fig:Menv-Lbol_zoom}, we predict a final median mass of $\overline{M_{\star}^{\rm{final}}} \sim$ 0.4\,$M_{\odot}$. 
Moreover, the most massive Class~0 object from our sample (SerpS-MM18) has $M_{\rm{env}}^{\rm{max}} \sim$3$\pm$1 $M_{\odot}$, and evolutionary models therefore suggest a maximum final stellar mass $M_{\star,max}^{\rm{final}} \sim$1.7$\pm$0.5 $M_{\odot}$. 
\\Thus, the use of the model evolutionary tracks drawn in Fig.\,\ref{fig:Menv-Lbol_zoom} predicts both a median and maximum stellar mass in agreement with a random sampling from the standard IMF \citep{Kroupa01,Chabrier03a}. 

\subsection{Two modes of efficient clustered star formation?} 

The high values of SFRs we derive for both Serpens South and W40 are consistent with the fact that we consider dense and compact regions actively forming large numbers of protostars from large mass reservoirs, i.e protoclusters 
\citep[cf.][]{Hurt96, Myers01, Andre00, Lada03}. 
These two adjacent star-forming regions are likely to share the same distance and have similar projected areas with ongoing star formation  as well as similar gas reservoirs 
(gas masses of 610~$M_{\odot}$ and 310~$M_{\odot}$, respectively).
However, the formation of each of these two young protostellar populations is likely to have been initiated in a different way:

\noindent (1)~Large-scale collapse of a gravitationally unstable filament for the Serpens South cluster-forming clump, similar to what was observed in the intermediate-mass protocluster NGC~2264-C \citep{Peretto06}. 
Indeed, it has been proposed based on the first results from \herschel\  \citep{Andre10} that a dense filament with a highly supercritical mass per unit length \citep[cf.][]{Inutsuka97} such as Serpens South may become globally gravitationally unstable and undergo overall collapse, eventually leading to the formation of a rich protocluster.
Our MAMBO 1.2~mm map confirm that the high-column-density filamentary clump surrounding the central mid-infrared embedded cluster detected 
by \spitzer\ \citep{Gutermuth08} in Serpens-South is actively forming young Class~0 protostars. 
The high fraction of Class~0 and Class~I protostars compared to the total number (30--40) of Class~II YSOs, found in the central core of the protocluster but also in its filamentary dusty arms, 
is suggestive of a recent burst of star formation in this cluster-forming clump. 
This view is also supported by the recent results of \citet{Nakamura11b}, who detected numerous protostellar outflows toward this clump.

\noindent (2)~Pressure-triggered star cluster formation (see \citealt{Elmegreen98,Deharveng05}) around the H{\small II} region W40. 
The winds and energetic radiation from the massive young stars of W40 have created an expanding H{\small II} region 
and compressed the gas clump at the origin of the cluster, forming a ring-like structure of dense gas 
now harboring a second generation of young protostars. 
Indeed, all of the youngest (Class~0) YSOs are found in the high column-density regions surrounding the infrared cluster containing most of the more 
evolved Class~II and Class~III objects, while not many of these more evolved objects are found in the outer dense regions harboring the Class~0 protostars. 
If, as suggested by the high disk fraction reported by \citet{Kuhn10}, the young infrared cluster is indeed $\sim$1~Myr old, then W40 is currently undergoing a small burst of star formation, and may end up forming a rich bound cluster.

We have shown that these two protoclusters, although resulting from different collapse initial conditions, are presently harboring similar numbers of Class~0 protostars. 
This suggests that after the onset of gravitational collapse, the detailed manner in which the collapse has been initiated does not affect much the ability of the clump to form stars.
Moreover, both W40 and Serpens South are characterized by large SFRs,  an order of magnitude higher than the typical SFRs observed for embedded infrared clusters \citep{Lada03}. 
This suggests that (sub)millimeter protoclusters are star forming clumps caught in the very act of efficiently forming stars, while embedded infrared clusters 
may already be in the decay phase following the initial burst of star formation.
If they last long enough, such high SFRs are likely to produce a large number of stars before gas dispersal processes (outflows, radiation, etc.) from the growing cluster membership terminate the process (see the conclusions of \citealt{Nakamura11b} regarding Serpens South), i.e reach SFEs~$>$30\% which would allow them to remain bound \citep{Lada03, Baumgardt07, Goodwin09}. 
W40 and Serpens South are therefore different but good candidates to form rich bound clusters.

~\\

\noindent {\it{\small{Acknowledgments:}}} {\small{We are grateful to the specialist astronomy group dedicated to star formation within the SPIRE Consortium ("SAG 3")
for letting us use a preliminary version of their {\it{Herschel}} photometric results in advance of publication. A first-generation paper including the full {\it{Herschel}} sources catalog in the Aquila Rift complex will be published shortly (K\"onyves et al., in prep.).\\
We thank R. Gutermuth for providing us with its list of \spitzer\ IRAC Class~I sources in the Serpens South field prior to publication.\\
We thank M. Kuhn, E. Feigelson and K. Getman for useful comments on the $Chandra$ PMS population in W40.}}

 \bibliographystyle{aa}
    %\bibliography{bibliographie}

\begin{thebibliography}{75}
\expandafter\ifx\csname natexlab\endcsname\relax\def\natexlab#1{#1}\fi

\bibitem[{{Altenhoff} {et~al.}(1970){Altenhoff}, {Downes}, {Goad}, {Maxwell},
  \& {Rinehart}}]{Altenhoff70}
{Altenhoff}, W.~J., {Downes}, D., {Goad}, L., {Maxwell}, A., \& {Rinehart}, R.
  1970, \aaps, 1, 319

\bibitem[{{Alves} {et~al.}(2007){Alves}, {Lombardi}, \& {Lada}}]{Alves07}
{Alves}, J., {Lombardi}, M., \& {Lada}, C.~J. 2007, \aap, 462, L17

\bibitem[{{Andr{\'e}} {et~al.}(2007){Andr{\'e}}, {Belloche}, {Motte}, \&
  {Peretto}}]{Andre07}
{Andr{\'e}}, P., {Belloche}, A., {Motte}, F., \& {Peretto}, N. 2007, \aap, 472,
  519

\bibitem[{{Andr{\'e}} {et~al.}(2010){Andr{\'e}}, {Men'shchikov}, {Bontemps},
  {K{\"o}nyves}, {Motte}, {Schneider}, {Didelon}, {Minier}, {Saraceno},
  {Ward-Thompson}, {di Francesco}, {White}, {Molinari}, {Testi}, {Abergel},
  {Griffin}, {Henning}, {Royer}, {Mer{\'{\i}}n}, {Vavrek}, {Attard},
  {Arzoumanian}, {Wilson}, {Ade}, {Aussel}, {Baluteau}, {Benedettini},
  {Bernard}, {Blommaert}, {Cambr{\'e}sy}, {Cox}, {di Giorgio}, {Hargrave},
  {Hennemann}, {Huang}, {Kirk}, {Krause}, {Launhardt}, {Leeks}, {Le Pennec},
  {Li}, {Martin}, {Maury}, {Olofsson}, {Omont}, {Peretto}, {Pezzuto}, {Prusti},
  {Roussel}, {Russeil}, {Sauvage}, {Sibthorpe}, {Sicilia-Aguilar}, {Spinoglio},
  {Waelkens}, {Woodcraft}, \& {Zavagno}}]{Andre10}
{Andr{\'e}}, P., {Men'shchikov}, A., {Bontemps}, S., {et~al.} 2010, \aap, 518,
  L102+

\bibitem[{{Andr{\'e}} {et~al.}(2008){Andr{\'e}}, {Minier}, {Gallais},
  {Rev{\'e}ret}, {Le Pennec}, {Rodriguez}, {Boulade}, {Doumayrou}, {Dubreuil},
  {Lortholary}, {Martignac}, {Talvard}, {De Breuck}, {Hamon}, {Schneider},
  {Bontemps}, {Lagage}, {Pantin}, {Roussel}, {Miller}, {Purcell}, {Hill}, \&
  {Stutzki}}]{Andre08b}
{Andr{\'e}}, P., {Minier}, V., {Gallais}, P., {et~al.} 2008, \aap, 490, L27

\bibitem[{{Andr\'e} \& {Montmerle}(1994)}]{Andre94a}
{Andr\'e}, P. \& {Montmerle}, T. 1994, \apj, 420, 837

\bibitem[{{Andr\'e} {et~al.}(1993){Andr\'e}, {Ward-Thompson}, \&
  {Barsony}}]{Andre93}
{Andr\'e}, P., {Ward-Thompson}, D., \& {Barsony}, M. 1993, \apj, 406, 122

\bibitem[{{Andr\'e} {et~al.}(2000){Andr\'e}, {Ward-Thompson}, \&
  {Barsony}}]{Andre00}
{Andr\'e}, P., {Ward-Thompson}, D., \& {Barsony}, M. 2000, Protostars and
  Planets IV, 59

\bibitem[{{Baumgardt} \& {Kroupa}(2007)}]{Baumgardt07}
{Baumgardt}, H. \& {Kroupa}, P. 2007, \mnras, 380, 1589

\bibitem[{{Belloche} {et~al.}(2011){Belloche}, {Schuller}, {Parise},
  {Andr{\'e}}, {Hatchell}, {J{\o}rgensen}, {Bontemps}, {Wei{\ss}}, {Menten}, \&
  {Muders}}]{Belloche11}
{Belloche}, A., {Schuller}, F., {Parise}, B., {et~al.} 2011, \aap, 527, A145+

\bibitem[{{Bontemps} {et~al.}(2010){Bontemps}, {Andr{\'e}}, {K{\"o}nyves},
  {Men'shchikov}, {Schneider}, {Maury}, {Peretto}, {Arzoumanian}, {Attard},
  {Motte}, {Minier}, {Didelon}, {Saraceno}, {Abergel}, {Baluteau}, {Bernard},
  {Cambr{\'e}sy}, {Cox}, {di Francesco}, {di Giorgo}, {Griffin}, {Hargrave},
  {Huang}, {Kirk}, {Li}, {Martin}, {Mer{\'{\i}}n}, {Molinari}, {Olofsson},
  {Pezzuto}, {Prusti}, {Roussel}, {Russeil}, {Sauvage}, {Sibthorpe},
  {Spinoglio}, {Testi}, {Vavrek}, {Ward-Thompson}, {White}, {Wilson},
  {Woodcraft}, \& {Zavagno}}]{Bontemps10}
{Bontemps}, S., {Andr{\'e}}, P., {K{\"o}nyves}, V., {et~al.} 2010, \aap, 518,
  L85+

\bibitem[{{Bontemps} {et~al.}(1996){Bontemps}, {Andr{\'e}}, {Terebey}, \&
  {Cabrit}}]{Bontemps96a}
{Bontemps}, S., {Andr{\'e}}, P., {Terebey}, S., \& {Cabrit}, S. 1996, \aap,
  311, 858

\bibitem[{{Bressert} {et~al.}(2010){Bressert}, {Bastian}, {Gutermuth},
  {Megeath}, {Allen}, {Evans}, {Rebull}, {Hatchell}, {Johnstone}, {Bourke},
  {Cieza}, {Harvey}, {Merin}, {Ray}, \& {Tothill}}]{Bressert10}
{Bressert}, E., {Bastian}, N., {Gutermuth}, R., {et~al.} 2010, \mnras, 409, L54

\bibitem[{{Carpenter}(2000)}]{Carpenter00}
{Carpenter}, J.~M. 2000, \aj, 120, 3139

\bibitem[{{Chabrier}(2003)}]{Chabrier03a}
{Chabrier}, G. 2003, \pasp, 115, 763

\bibitem[{{Chen} {et~al.}(1995){Chen}, {Myers}, {Ladd}, \& {Wood}}]{Chen95}
{Chen}, H., {Myers}, P., {Ladd}, E., \& {Wood}, D. 1995, \apj, 445, 377

\bibitem[{{Deharveng} {et~al.}(2005){Deharveng}, {Zavagno}, \&
  {Caplan}}]{Deharveng05}
{Deharveng}, L., {Zavagno}, A., \& {Caplan}, J. 2005, \aap, 433, 565

\bibitem[{{Dzib} {et~al.}(2010){Dzib}, {Loinard}, {Mioduszewski}, {Boden},
  {Rodr{\'{\i}}guez}, \& {Torres}}]{Dzib10}
{Dzib}, S., {Loinard}, L., {Mioduszewski}, A.~J., {et~al.} 2010, \apj, 718, 610

\bibitem[{{Elmegreen}(1998)}]{Elmegreen98}
{Elmegreen}, B.~G. 1998, in Astronomical Society of the Pacific Conference
  Series, Vol. 148, Origins, ed. {C.~E.~Woodward, J.~M.~Shull, \&
  H.~A.~Thronson Jr.}, 150--+

\bibitem[{{Enoch} {et~al.}(2009){Enoch}, {Evans}, {Sargent}, \&
  {Glenn}}]{Enoch09}
{Enoch}, M.~L., {Evans}, N.~J., {Sargent}, A.~I., \& {Glenn}, J. 2009, \apj,
  692, 973

\bibitem[{{Enoch} {et~al.}(2007){Enoch}, {Glenn}, {Evans}, {Sargent}, {Young},
  \& {Huard}}]{Enoch07}
{Enoch}, M.~L., {Glenn}, J., {Evans}, II, N.~J., {et~al.} 2007, \apj, 666, 982

\bibitem[{{Evans} {et~al.}(2009){Evans}, {Dunham}, {J{\o}rgensen}, {Enoch},
  {Mer{\'{\i}}n}, {van Dishoeck}, {Alcal{\'a}}, {Myers}, {Stapelfeldt},
  {Huard}, {Allen}, {Harvey}, {van Kempen}, {Blake}, {Koerner}, {Mundy},
  {Padgett}, \& {Sargent}}]{Evans09}
{Evans}, N.~J., {Dunham}, M.~M., {J{\o}rgensen}, J.~K., {et~al.} 2009, \apjs,
  181, 321

\bibitem[{{Ginsburg} {et~al.}(2009){Ginsburg}, {Bally}, {Yan}, \&
  {Williams}}]{Ginsburg09}
{Ginsburg}, A.~G., {Bally}, J., {Yan}, C., \& {Williams}, J.~P. 2009, \apj,
  707, 310

\bibitem[{{Goodwin}(2009)}]{Goodwin09}
{Goodwin}, S.~P. 2009, \apss, 324, 259

\bibitem[{{Greene} {et~al.}(1994){Greene}, {Wilking}, {Andr\'e}, {Young}, \&
  {Lada}}]{Greene94}
{Greene}, T., {Wilking}, B., {Andr\'e}, P., {Young}, E., \& {Lada}, C. 1994,
  \apj, 434, 614

\bibitem[{{Gutermuth} {et~al.}(2008){Gutermuth}, {Bourke}, {Allen}, {Myers},
  {Megeath}, {Matthews}, {J{\o}rgensen}, {Di Francesco}, {Ward-Thompson},
  {Huard}, {Brooke}, {Dunham}, {Cieza}, {Harvey}, \& {Chapman}}]{Gutermuth08}
{Gutermuth}, R.~A., {Bourke}, T.~L., {Allen}, L.~E., {et~al.} 2008, \apjl, 673,
  L151

\bibitem[{{Harvey} {et~al.}(2007){Harvey}, {Mer{\'{\i}}n}, {Huard}, {Rebull},
  {Chapman}, {Evans}, \& {Myers}}]{Harvey07}
{Harvey}, P., {Mer{\'{\i}}n}, B., {Huard}, T.~L., {et~al.} 2007, \apj, 663,
  1149

\bibitem[{{Henning} {et~al.}(1995){Henning}, {Michel}, \&
  {Stognienko}}]{Henning95}
{Henning}, T., {Michel}, B., \& {Stognienko}, R. 1995, \planss, 43, 1333

\bibitem[{{Henriksen} {et~al.}(1997){Henriksen}, {Andre}, \&
  {Bontemps}}]{Henriksen97}
{Henriksen}, R., {Andre}, P., \& {Bontemps}, S. 1997, \aap, 323, 549

\bibitem[{{Hillenbrand} \& {Carpenter}(2000)}]{Hillenbrand00}
{Hillenbrand}, L.~A. \& {Carpenter}, J.~M. 2000, \apj, 540, 236

\bibitem[{{Hurt} \& {Barsony}(1996)}]{Hurt96}
{Hurt}, R. \& {Barsony}, M. 1996, \apjl, 460, L45+

\bibitem[{{Inutsuka} \& {Miyama}(1997)}]{Inutsuka97}
{Inutsuka}, S. \& {Miyama}, S.~M. 1997, \apj, 480, 681

\bibitem[{{Kenyon} {et~al.}(1990){Kenyon}, {Hartmann}, {Strom}, \&
  {Strom}}]{Kenyon90}
{Kenyon}, S.~J., {Hartmann}, L.~W., {Strom}, K.~M., \& {Strom}, S.~E. 1990,
  \aj, 99, 869

\bibitem[{{Kirk} {et~al.}(2007){Kirk}, {Johnstone}, \& {Tafalla}}]{KirkH07}
{Kirk}, H., {Johnstone}, D., \& {Tafalla}, M. 2007, \apj, 668, 1042

\bibitem[{{Knee} \& {Sandell}(2000)}]{Knee00}
{Knee}, L.~B.~G. \& {Sandell}, G. 2000, \aap, 361, 671

\bibitem[{{Knude}(2011)}]{Knude11}
{Knude}, J. 2011, ArXiv:1103.0455

\bibitem[{{K{\"o}nyves} {et~al.}(2010){K{\"o}nyves}, {Andr{\'e}},
  {Men'shchikov}, {Schneider}, {Arzoumanian}, {Bontemps}, {Attard}, {Motte},
  {Didelon}, {Maury}, {Abergel}, {Ali}, {Baluteau}, {Bernard}, {Cambr{\'e}sy},
  {Cox}, {di Francesco}, {di Giorgio}, {Griffin}, {Hargrave}, {Huang}, {Kirk},
  {Li}, {Martin}, {Minier}, {Molinari}, {Olofsson}, {Pezzuto}, {Russeil},
  {Roussel}, {Saraceno}, {Sauvage}, {Sibthorpe}, {Spinoglio}, {Testi},
  {Ward-Thompson}, {White}, {Wilson}, {Woodcraft}, \& {Zavagno}}]{Konyves10}
{K{\"o}nyves}, V., {Andr{\'e}}, P., {Men'shchikov}, A., {et~al.} 2010, \aap,
  518, L106+

\bibitem[{{Kroupa}(2001)}]{Kroupa01}
{Kroupa}, P. 2001, \mnras, 322, 231

\bibitem[{{Kuhn} {et~al.}(2010){Kuhn}, {Getman}, {Feigelson}, {Reipurth},
  {Rodney}, \& {Garmire}}]{Kuhn10}
{Kuhn}, M.~A., {Getman}, K.~V., {Feigelson}, E.~D., {et~al.} 2010, \apj, 725,
  2485

\bibitem[{{Lada} \& {Lada}(2003)}]{Lada03}
{Lada}, C.~J. \& {Lada}, E.~A. 2003, \araa, 41, 57

\bibitem[{{Lada} {et~al.}(2010){Lada}, {Lombardi}, \& {Alves}}]{Lada10}
{Lada}, C.~J., {Lombardi}, M., \& {Alves}, J.~F. 2010, \apj, 724, 687

\bibitem[{{Loinard} {et~al.}(2011){Loinard}, {Mioduszewski}, {Torres}, {Dzib},
  {Rodriguez}, \& {Boden}}]{Loinard11}
{Loinard}, L., {Mioduszewski}, A.~J., {Torres}, R.~M., {et~al.} 2011, ArXiv
  e-prints

\bibitem[{{Matzner} \& {McKee}(2000)}]{Matzner00}
{Matzner}, C. \& {McKee}, C. 2000, \apj, 545, 364

\bibitem[{{Maury} {et~al.}(2009){Maury}, {Andr{\'e}}, \& {Li}}]{Maury09}
{Maury}, A., {Andr{\'e}}, P., \& {Li}, Z.-Y. 2009, \aap, 499, 175

\bibitem[{{Men'shchikov} {et~al.}(2011){Men'shchikov}, {Andr{\'e}}, {Didelon},
  {K{\"o}nyves}, {Schneider}, {Motte}, \& {Bontemps}}]{Menshchikov11}
{Men'shchikov}, A., {Andr{\'e}}, P., {Didelon}, P., {et~al.} 2011, in prep.

\bibitem[{{Men'shchikov} {et~al.}(2010){Men'shchikov}, {Andr{\'e}}, {Didelon},
  {K{\"o}nyves}, {Schneider}, {Motte}, {Bontemps}, {Arzoumanian}, {Attard},
  {Abergel}, {Baluteau}, {Bernard}, {Cambr{\'e}sy}, {Cox}, {di Francesco}, {di
  Giorgio}, {Griffin}, {Hargrave}, {Huang}, {Kirk}, {Li}, {Martin}, {Minier},
  {Miville-Desch{\^e}nes}, {Molinari}, {Olofsson}, {Pezzuto}, {Roussel},
  {Russeil}, {Saraceno}, {Sauvage}, {Sibthorpe}, {Spinoglio}, {Testi},
  {Ward-Thompson}, {White}, {Wilson}, {Woodcraft}, \&
  {Zavagno}}]{Menshchikov10}
{Men'shchikov}, A., {Andr{\'e}}, P., {Didelon}, P., {et~al.} 2010, \aap, 518,
  L103+

\bibitem[{{Meyer} {et~al.}(2000){Meyer}, {Adams}, {Hillenbrand}, {Carpenter},
  \& {Larson}}]{Meyer00}
{Meyer}, M.~R., {Adams}, F.~C., {Hillenbrand}, L.~A., {Carpenter}, J.~M., \&
  {Larson}, R.~B. 2000, Protostars and Planets IV, 121

\bibitem[{{Motte} {et~al.}(1998){Motte}, {Andr\'e}, \& {Neri}}]{Motte98}
{Motte}, F., {Andr\'e}, P., \& {Neri}, R. 1998, \aap, 336, 150

\bibitem[{{Myers} {et~al.}(1998){Myers}, {Adams}, {Chen}, \&
  {Schaff}}]{Myers98a}
{Myers}, P., {Adams}, F., {Chen}, H., \& {Schaff}, E. 1998, \apj, 492, 703

\bibitem[{{Myers}(2001)}]{Myers01}
{Myers}, P.~C. 2001, in Astronomical Society of the Pacific Conference Series,
  Vol. 243, From Darkness to Light: Origin and Evolution of Young Stellar
  Clusters, ed. {T.~Montmerle \& P.~Andr{\'e}}, 131--+

\bibitem[{{Myers} {et~al.}(1986){Myers}, {Dame}, {Thaddeus}, {Cohen},
  {Silverberg}, {Dwek}, \& {Hauser}}]{Myers86}
{Myers}, P.~C., {Dame}, T.~M., {Thaddeus}, P., {et~al.} 1986, \apj, 301, 398

\bibitem[{{Nakamura} {et~al.}(2011{\natexlab{a}}){Nakamura}, {Kamada},
  {Kamazaki}, {Kawabe}, {Kitamura}, {Shimajiri}, {Tsukagoshi}, {Tachihara},
  {Akashi}, {Azegami}, {Ikeda}, {Kurono}, {Li}, {Miura}, {Nishi}, \&
  {Umemoto}}]{Nakamura11a}
{Nakamura}, F., {Kamada}, Y., {Kamazaki}, T., {et~al.} 2011{\natexlab{a}},
  \apj, 726, 46

\bibitem[{{Nakamura} {et~al.}(2011{\natexlab{b}}){Nakamura}, {Sugitani},
  {Shimajiri}, {Tsukagoshi}, {Higuchi}, {Nishiyama}, {Kawabe}, {Takami},
  {Karr}, {Gutermuth}, \& {Wilson}}]{Nakamura11b}
{Nakamura}, F., {Sugitani}, K., {Shimajiri}, Y., {et~al.} 2011{\natexlab{b}},
  submitted to ApJ

\bibitem[{{Nutter} \& {Ward-Thompson}(2007)}]{Nutter07}
{Nutter}, D. \& {Ward-Thompson}, D. 2007, \mnras, 374, 1413

\bibitem[{{Olmi} \& {Testi}(2002)}]{Olmi02}
{Olmi}, L. \& {Testi}, L. 2002, \aap, 392, 1053

\bibitem[{{Ossenkopf} \& {Henning}(1994)}]{Ossenkopf94}
{Ossenkopf}, V. \& {Henning}, T. 1994, \aap, 291, 943

\bibitem[{{Palla} \& {Stahler}(1999)}]{Palla99}
{Palla}, F. \& {Stahler}, S.~W. 1999, \apj, 525, 772

\bibitem[{{Peretto} {et~al.}(2006){Peretto}, {Andr{\'e}}, \&
  {Belloche}}]{Peretto06}
{Peretto}, N., {Andr{\'e}}, P., \& {Belloche}, A. 2006, \aap, 445, 979

\bibitem[{{Peretto} {et~al.}(2007){Peretto}, {Hennebelle}, \&
  {Andr{\'e}}}]{Peretto07}
{Peretto}, N., {Hennebelle}, P., \& {Andr{\'e}}, P. 2007, \aap, 464, 983

\bibitem[{{Robitaille} {et~al.}(2007){Robitaille}, {Whitney}, {Indebetouw}, \&
  {Wood}}]{Robitaille07}
{Robitaille}, T.~P., {Whitney}, B.~A., {Indebetouw}, R., \& {Wood}, K. 2007,
  \apjs, 169, 328

\bibitem[{{Saraceno} {et~al.}(1996){Saraceno}, {Andre}, {Ceccarelli},
  {Griffin}, \& {Molinari}}]{Saraceno96}
{Saraceno}, P., {Andre}, P., {Ceccarelli}, C., {Griffin}, M., \& {Molinari}, S.
  1996, \aap, 309, 827

\bibitem[{{Schneider} {et~al.}(2011){Schneider}, {Bontemps}, {Simon},
  {Ossenkopf}, {Federrath}, {Klessen}, {Motte}, {Andr{\'e}}, {Stutzki}, \&
  {Brunt}}]{Schneider11}
{Schneider}, N., {Bontemps}, S., {Simon}, R., {et~al.} 2011, \aap, 529, A1+

\bibitem[{{Smith} {et~al.}(1985){Smith}, {Bentley}, {Castelaz}, {Gehrz},
  {Grasdalen}, \& {Hackwell}}]{Smith85}
{Smith}, J., {Bentley}, A., {Castelaz}, M., {et~al.} 1985, \apj, 291, 571

\bibitem[{{Smith} {et~al.}(2011){Smith}, {Fellhauer}, {Goodwin}, \&
  {Assmann}}]{SmithR11a}
{Smith}, R., {Fellhauer}, M., {Goodwin}, S., \& {Assmann}, P. 2011, \mnras, 601

\bibitem[{{Stahler}(1988)}]{Stahler88}
{Stahler}, S. 1988, \apj, 332, 804

\bibitem[{{Strai{\v z}ys} {et~al.}(1996){Strai{\v z}ys}, {{\v C}ernis}, \&
  {Barta{\v s}i{\= u}t{\.e}}}]{Straizys96}
{Strai{\v z}ys}, V., {{\v C}ernis}, K., \& {Barta{\v s}i{\= u}t{\.e}}, S. 1996,
  Baltic Astronomy, 5, 125

\bibitem[{{Strai{\v z}ys} {et~al.}(2003){Strai{\v z}ys}, {{\v C}ernis}, \&
  {Barta{\v s}i{\= u}t{\.e}}}]{Straizys03}
{Strai{\v z}ys}, V., {{\v C}ernis}, K., \& {Barta{\v s}i{\= u}t{\.e}}, S. 2003,
  \aap, 405, 585

\bibitem[{{Testi} {et~al.}(1999){Testi}, {Palla}, \& {Natta}}]{Testi99}
{Testi}, L., {Palla}, F., \& {Natta}, A. 1999, \aap, 342, 515

\bibitem[{{Testi} \& {Sargent}(1998)}]{Testi98}
{Testi}, L. \& {Sargent}, A. 1998, \apjl, 508, L91

\bibitem[{{Vallee}(1987)}]{Vallee87}
{Vallee}, J.~P. 1987, \aap, 178, 237

\bibitem[{{Vallee} {et~al.}(1992){Vallee}, {Guilloteau}, \&
  {MacLeod}}]{Vallee92}
{Vallee}, J.~P., {Guilloteau}, S., \& {MacLeod}, J.~M. 1992, \aap, 266, 520

\bibitem[{{Visser} {et~al.}(2002){Visser}, {Richer}, \& {Chandler}}]{Visser02}
{Visser}, A., {Richer}, J., \& {Chandler}, C. 2002, \aj, 124, 2756

\bibitem[{{Wilking} \& {Lada}(1983)}]{Wilking83}
{Wilking}, B. \& {Lada}, C. 1983, \apj, 274, 698

\bibitem[{{Young} \& {Evans}(2005)}]{Young05}
{Young}, C.~H. \& {Evans}, II, N.~J. 2005, \apj, 627, 293

\bibitem[{{Zhu} {et~al.}(2006){Zhu}, {Wu}, \& {Wei}}]{Zhu06}
{Zhu}, L., {Wu}, Y., \& {Wei}, Y. 2006, \cjaa, 6, 61

\end{thebibliography}

%%%%%%
\begin{figure*}[!t]
\centering
\includegraphics[width=0.07\textwidth,angle=0,trim=0cm 0cm 0cm 0cm,clip=true]{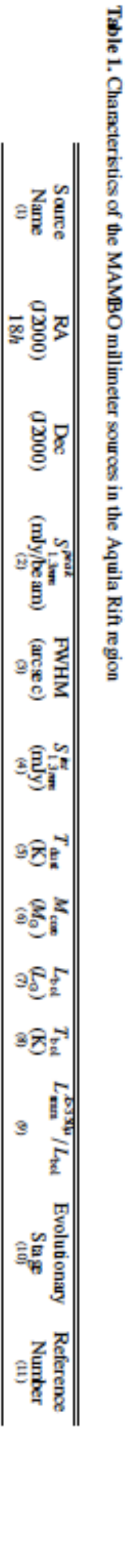}
\label{tab:tabp1}
\caption{The table listing the properties of the MAMBO millimeter sources in the Aquila Rift region is available upon 
request to the first author.}
\end{figure*}
%%%%%%
%%%%%%
%\begin{figure*}[!t]
%\centering
%\includegraphics[width=0.98\textwidth,angle=0,trim=0cm 0cm 0cm 0cm,clip=true]{Table_p2.pdf}
%\label{tab:tabp2}
%\end{figure*}
%%%%%%

\end{document}